\documentclass[letterpaper,aps,twocolumn,prl,amsmath,amssymb,floatfix,showpacs]{revtex4}
\usepackage{amsmath,amssymb,amsfonts,mathrsfs}
\usepackage{color}
\usepackage{epsfig}
\usepackage{psfrag}
\usepackage{graphicx}
\usepackage{color}

\begin{document}

\title{Electrical and Thermal Transport in Inhomogeneous Luttinger Liquids}

\author{Wade DeGottardi and K. A. Matveev}
\affiliation{\small{Materials Science Division, Argonne National Laboratory, Argonne, Illinois 60439, USA }}

\date{\today}

\begin{abstract}
We study the transport properties of long quantum wires by generalizing the Luttinger liquid approach to allow for the finite lifetime of the bosonic excitations. Our theory accounts for long-range disorder and strong electron interactions, both of which are common features of experiments with quantum wires. We obtain the electrical and thermal resistances and thermoelectric properties of such quantum wires. We cast our results in terms of the thermal conductivity and bulk viscosity of the electron liquid and give the temperature scale above which the transport can be described by classical hydrodynamics.
\end{abstract}


\maketitle

\emph{Introduction.}-- The quantization of conductance exhibited by quantum wires in multiples of $G_0 = 2 e^2 / h$ highlights the crucial role played by the confinement of the electronic wave function in directions transverse to the wire~\cite{vanwees,pepper,transport,giamarchi}. Quantum wires potentially represent a window into the rich array of non-Fermi liquid phenomena predicted for one-dimensional electron systems~\cite{giamarchi,haldane}. The most direct means of probing such wires experimentally is through their transport properties. In addition to reduced dimensionality, aspects of these system which are crucial to understanding these properties include strong electron-electron interactions as well as slowly varying disorder potentials which inevitably arise in the fabrication of such devices. While fabrication techniques are now able to limit sources of short range disorder, such as impurities in heterostructure realizations, long range disorder remains a feature of these systems as it arises from the process of modulation doping used to populate the two-dimensional electron gas from which quantum wires are patterned~\cite{shortrange}.

At sufficiently high temperatures, the transport properties of electron liquids can be described by classical hydrodynamics~\cite{hydro}. This approach offers an ostensibly classical description of systems in cases for which the characteristic length scale associated with violations of momentum conservation is much longer than the electron-electron mean-free path. In one dimension, the conductance can be expressed in terms of the thermal conductivity $\kappa$ and the bulk viscosity $\zeta$ of the electron liquid~\cite{hydro}.

Ultimately, a complete theoretical description of these wires requires a quantum mechanical treatment. The case of weakly interacting electrons in a wire with long range disorder has been studied~\cite{levchenko}. However, the applicability of this theory to experiments is limited since electron-electron interactions in wires are typically strong. A commonly employed theoretical framework for studying interacting electrons in one dimension is the Luttinger liquid (LL) formalism, a powerful non-perturbative approach which can account for electron interactions of arbitrary strength. In this framework, the excitations of the electron liquid are described by non-interacting bosons~\cite{giamarchi,haldane}. The effects of short range disorder on the conductance properties of a LL are well understood~\cite{schulz,kanefisher,furusaki}, however the effects of long range inhomogeneities have not been explored.

\begin{figure}[b]
\begin{center}
    \includegraphics[width = 0.45\textwidth]{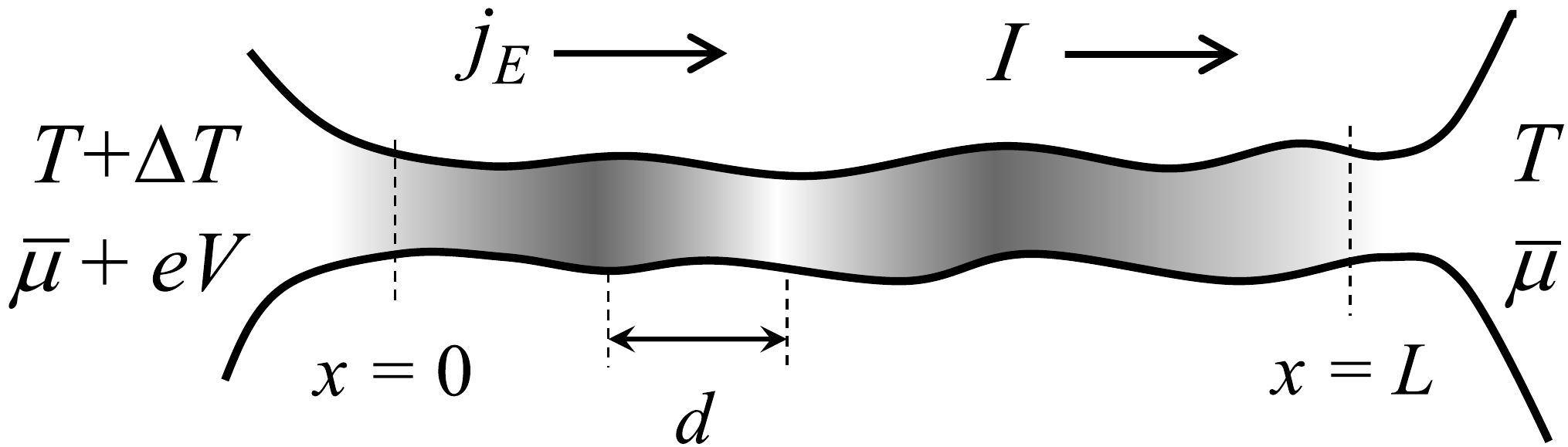}
     \caption{Sketch of an inhomogeneous quantum wire in contact with Fermi liquid leads with temperatures $T$ and $T + \Delta T$ and electrochemical potentials $\bar{\mu}$ and $\bar{\mu}+eV$. The arrows indicate the flow of energy and electric currents, denoted by $j_E$ and $I$, respectively. The shading in the wire indicates the non-uniform electron density, $n(x)$. The length scale $d$ characterizes typical variations of the electron density.}
     \label{fig:setup}
\end{center}
\end{figure}

In this work, we study an inhomogeneous LL and calculate its transport properties. Standard LL theory, in which the bosonic excitations are infinitely long-lived, predicts that the conductance of quantum wire remains $G_0$ even for strongly interacting electrons~\cite{maslov}. Accounting for the scattering of the bosonic excitations, we find that the conductance is suppressed below $G_0$. In addition to the electrical resistance, we obtain expressions for the thermal resistance, the Peltier coefficient, and the thermopower.

Our theory holds insofar as LL theory is applicable. In particular, the temperature $T$ must be lower than the bandwidth $D$, which is typically on the order of the Fermi energy of the electron liquid. We identify a temperature scale $T^\ast$ such that for $T^\ast \ll T \ll D$, our result for the resistance of a quantum wire with weak disorder reduces to that of the hydrodynamical theory~\cite{hydro}. An interesting aspect of our results is that although they differ from those of classical hydrodynamics at low temperatures, $T \lesssim T^\ast$, they can still be expressed in terms of the thermal conductivity $\kappa$ and bulk viscosity $\zeta$ of the electron liquid.

The system of interest is shown in Fig.~1. For clarity and simplicity of presentation, we focus on the case of a spinless electron liquid, and briefly discuss the generalization to spinful and multi-channel LLs as well. We consider an inhomogeneous LL in which the electron density $n(x)$ has spatial variations of characteristic length $d \gg n^{-1}$. On length scales much smaller than $d$, segments of the wire can be treated as uniform LLs.

\emph{Uniform Luttinger liquid.}-- In the LL approximation, particle-hole excitations of the electron system are described by non-interacting bosons. The number of right- and left-moving electrons (denoted by $N_{R/L}$) is held fixed. The Hamiltonian and the momentum of the LL are given by
\begin{eqnarray}
H &=& \sum_p \varepsilon_p b_p^\dagger b_p^{\phantom\dagger} + \frac{\pi \hbar}{2 L} \left[ v_N \left( N - N_0 \right)^2 + v_J J^2 \right], \label{eq:Ham} \\
P &=& \sum_p p b_p^\dagger b_p^{\phantom\dagger} +  p_F J,
\label{eq:momentum}
\end{eqnarray}
where $N = N_R + N_L$, $J = N_R - N_L$, and $b_p^\dagger$ creates a bosonic excitation of momentum $p$ and energy $\varepsilon_p$. The Fermi momentum is proportional to the electron density, i.e. $p_F = \pi \hbar n$. The quantity $N_0$ represents a fiducial number of electrons in the system. The velocities $v_N$ and $v_J$ are renormalized from the Fermi velocity by the interactions~\cite{haldane}.

Standard LL theory holds that the bosons have an acoustic spectrum $\varepsilon_p = v|p|$, where $v$ depends on the interaction strength. While this description is appropriate for a liquid at rest, the spectrum is altered by bulk motion of the fluid. For a fluid moving with velocity $v_d$, Galilean invariance gives that the spectrum in the lab frame is $\varepsilon_p = v|p| + v_d p$. This is consistent with the fact that in the lab frame, the right- and left-moving excitations have velocities $\pm v + v_d$, respectively.

\emph{Equilibration processes.}-- In a uniform LL, there are two types of relaxation processes which occur with widely separated timescales. Fast collisions among the bosons, which conserve their momentum, lead to a partially equilibrated distribution function
\begin{equation}
N_p = \frac{1}{e^{(\varepsilon_p - up)/T}-1},
\label{eq:distribution}
\end{equation}
where the parameter $u$ fixes the value of the total momentum of the bosons. Physically, $u$ is the velocity of the gas of the bosonic excitations. The rate at which bosons scatter, denoted here by $\tau_0^{-1}$, is expected to scale as a power of temperature~\cite{revmodphys,ristivojevic,lin, gangardt}.

Equation~(\ref{eq:distribution}) does not describe a fully equilibrated liquid. At full equilibrium, the velocity of the bosons $u$ equals the velocity of the electronic fluid $v_d$~\cite{spinless}. For $u \neq v_d$, the velocity $u$ can relax to $v_d$ by processes in which momentum is exchanged between the bosonic excitations and the zero mode $J$ [see Eq.~(\ref{eq:momentum})] and this relaxation is described by
\begin{equation}
\dot{u} = -\frac{u - v_d}{\tau}.
\label{eq:dotu}
\end{equation}
The mechanism of relaxation of involves the backscattering of an electron from one Fermi point to the other and a transfer of $2p_F$ of momentum to the bosons as dictated by the conservation of the total momentum $P$ [Eq.~(\ref{eq:momentum})]. This process requires that a hole pass through the bottom of the band and thus the rate has an Arrhenius activated form $\tau^{-1} \propto e^{-D/T}$, where the activation energy is on the order of the bandwidth of the system~\cite{spinless}. A detailed microscopic calculation of the quantity $\tau$ is given in~\cite{equilibration}.

The following analysis applies in the regime $T \ll D$ for which $\tau \gg \tau_0$. Our calculation of the transport properties of the LL involves the evaluation of the temperature gradient by tracking the momentum of the gas of excitations.  We work in the linear response regime and thus retain any terms linear in $u$ or $v_d$. The momentum density of bosons with occupation numbers $N_p$ given by Eq.~(\ref{eq:distribution}) is
\begin{equation}
\rho_P = \int_{-\infty}^\infty \frac{dp}{h} p N_p = \frac{\pi T^2}{3 \hbar v^3} \left( u-v_d \right).
\label{eq:rhop}
\end{equation}
The momentum current of the gas of excitations is
\begin{equation}
j_P = \int_{-\infty}^\infty \frac{dp}{h} v_p p N_p = \frac{\pi T^2}{6 \hbar v},
\label{eq:jp}
\end{equation}
where $v_p = \mbox{sgn}(p) v + v_d$. We will also make use of the energy density and current which are given by $\rho_E = j_P$ and $j_E = T s_0 n u$, expressions valid to linear order in $u$ and $v_d$. Here, $s_0$ is the entropy per particle of a fluid in full equilibrium, $s_0 = \pi T / 3 \hbar n v$.

Taking a time derivative of Eq.~(\ref{eq:rhop}) yields
\begin{equation}
\dot{\rho}_{P}^{(\kappa)} = - \frac{\pi T^2}{3 \hbar v^3} \, \frac{u - v_d}{\tau} = - \frac{s_0 n}{\kappa} \left(j_E - \frac{T s_0 I}{e} \right),
\label{eq:rhop1}
\end{equation}
where we have used the expression (\ref{eq:dotu}) for $\dot{u}$ and the fact that $\dot{v}_d = 0$. The quantity $\kappa = \pi T v \tau / 3 \hbar$ has been introduced; we will see below that $\kappa$ is the thermal conductivity of the electron liquid. Because relaxation is due to the backscattering of electrons, conservation of the total momentum (\ref{eq:momentum}) requires that
\begin{equation}
\dot{n}^R = - \dot{\rho}_{P}^{(\kappa)} / 2 p_F,
\label{eq:dotnr}
\end{equation}
where $\dot{n}^R$ is the change in the number of right-movers per unit length, i.e. $\dot{N}^R = \int dx \ \dot{n}^R$.

\emph{Inhomogeneous LL.}-- We now consider the effect of inhomogeneities of the LL characterized by the spatial variations of the electron density $n(x)$. We assume that the scale of the inhomogeneities is large, i.e. $d \gg v \tau_0$. In this limit, the boson distribution function in the wire is described by Eq.~(\ref{eq:distribution}) with spatially varying $u$ and $T$. While $\dot{\rho}_{P}^{(\kappa)}$ involves a redistribution of the momentum between the bosons and the zero modes, inhomogeneities lead to a loss of the total momentum of the liquid.

Due to the fact that the velocity $v$ is controlled by the electron density, the spatial dependence of $n(x)$ leads to spatial variations in the velocity. Energy conservation dictates that as the velocity  $v(x)$ of a ballistically propagating boson changes, so does its momentum. In order to calculate the resultant contribution to $\dot{\rho}_P$,  we consider a semiclassical description of the bosons in which their energy $\varepsilon_p(x)$ is a function of momentum and position. Hamilton's equations of motion give, $\dot{p} = - \partial_x \varepsilon_p = - \partial_x v |p| - ( \partial_x v_d ) p$. This gives rise to a change of the momentum density of the gas of excitations
\begin{equation}
\dot{\rho}_{P}^{(0)} = \int \frac{dp}{h} \ \dot{p} N_p = - \frac{\pi T^2}{6 \hbar v^2} \partial_x v,
\label{eq:rhop3}
\end{equation}
with no terms linear in $u$ or $v_d$ appearing.

In addition to changing the momentum of ballistically propagating bosons, inhomogeneities give rise to scattering processes which also result in a contribution to $\dot{\rho}_P$. In these dissipative processes, right-moving bosons scatter off inhomogeneities and become left-movers~\cite{rech}. The reverse process also occurs, though not at the same rate. Given the linearity of the bosonic spectrum, $u$ in Eq.~(\ref{eq:distribution}) can be interpreted as giving rise to a difference in the effective temperatures of the right-($R$) and left-($L$) moving bosons, i.e. $T_{R/L} = T / \left[ 1 - u / ( \pm v + v_d ) \right]$. As a result of these scattering events, energy flows from the warmer subsystem ($R$) to the cooler one ($L$) as right-moving bosons are converted to left-movers. For $u = 0$ ($T_R = T_L$), the two branches are in equilibrium. This implies that for $u \ll v$, the rate at which momentum is lost is proportional to $u$. Since these processes are driven by inhomogeneities, the scattering amplitude must be proportional to $d^{-1} \sim \partial_x n/n$, i.e. the inverse length scale characterizing the disorder, while the corresponding rate is proportional to the square of this quantity. Therefore, we obtain
\begin{equation}
\dot{\rho}_{P}^{(\zeta)} = - \zeta \left( \frac{\partial_x n}{n} \right)^2 u,
\label{eq:rhop2}
\end{equation}
where we have introduced a parameter $\zeta$ with units of momentum. That $\zeta$ is indeed the bulk viscosity will be demonstrated in our discussion of the transport properties. In contrast to Eq.~(\ref{eq:rhop1}) which describes the exchange of momentum between the bosons and the zero modes, Eq.~(\ref{eq:rhop2}) represents a net loss of the momentum of the full electron liquid as bosons scatter off inhomogeneities.

The arguments leading to Eq.~(\ref{eq:rhop2}) are quite general and this result holds for multi-channel Luttinger liquids. For the particular case of a fully equilibrated, single channel LL, $\dot{\rho}_{P}^{(\zeta)}$ was evaluated in Ref.~\cite{rech}. The result is consistent with Eq.~(\ref{eq:rhop2}) provided
\begin{equation}
\zeta = \frac{T v}{4} \left( \partial_n \frac{n}{v} \right)^2.
\label{eq:vis}
\end{equation}
The bulk viscosity captures the response of a fluid to changes in its density and thus it is quite natural that Eq.~(\ref{eq:vis}) involves a derivative with respect to $n$.

Equation~(\ref{eq:rhop2}) describes damping which brings the gas of excitations to rest in the lab frame, i.e. $u \rightarrow 0$. These processes counteract those described by Eq.~(\ref{eq:rhop1}) which drive $u \rightarrow v_d$. As is true for the case of weakly interacting electrons~\cite{levchenko}, the establishment of the velocity $u(x)$ results from the competition between the effects of electron backscattering and the scattering of excitations by spatial inhomogeneities.

\emph{Transport properties.}-- The derivation of the transport coefficients requires that we relate $\Delta T$ and $V$ to the energy and electric currents, $j_E$ and $I$ (see Fig. 1). The temperature gradient can be obtained by tracking the momentum of the gas of excitations. In the steady-state regime, $\partial_t \rho_P = 0$, and thus the gradient of momentum current obeys
\begin{equation}
\partial_x j_P = \dot{\rho}_P^{(0)} + \dot{\rho}_P^{(\kappa)} + \dot{\rho}_P^{(\zeta)}.
\label{eq:momentumbosons}
\end{equation}
In thermal equilibrium, $u = v_d = 0$ and Eqs.~(\ref{eq:rhop1}) and (\ref{eq:rhop2}) require that $\dot{\rho}_P^{(\kappa)}$ and $\dot{\rho}_P^{(\zeta)}$ vanish. Thus, we have $\partial_x j_P = \dot{\rho}_P^{(0)}$. Indeed, Eqs.~(\ref{eq:jp}) and (\ref{eq:rhop3}) satisfy this relation for $\partial_x T = 0$, as is necessarily the case for a system in thermal equilibrium.

For a system out of equilibrium, Eqs.~(\ref{eq:rhop1}) and (\ref{eq:rhop2}) show that $\dot{\rho}_P^{(\kappa)}$ and $\dot{\rho}_P^{(\zeta)}$ contribute terms linear in $j_E = T s_0 n u$ and $I = e n v_d$ to the right-hand side of Eq.~(\ref{eq:momentumbosons}). However, $j_P$ has no linear correction in these quantities. The only way that Eq.~(\ref{eq:momentumbosons}) can be satisified is for the temperature to acquire a spatial gradient. Substituting $j_P$ [Eq.~(\ref{eq:jp})] into Eq.~(\ref{eq:momentumbosons}) reveals that the temperature gradient of the LL is
\begin{equation}
\partial_x T = - \frac{j_E}{\kappa} + \frac{T s_0 I}{\kappa e} - \frac{\zeta}{T s_0^2} \left( \partial_x \frac{1}{n} \right)^2 j_E.
\label{eq:thermgrad}
\end{equation}
For $I = 0$, Eq.~(\ref{eq:thermgrad}) establishes that $\kappa$, as defined after Eq.~(\ref{eq:rhop1}), is the thermal conductivity of a uniform electron liquid. The last term in Eq.~(\ref{eq:thermgrad}) shows that spatial variations in $n(x)$ give a correction to the thermal resistivity of the system.

We now demonstrate that the quantity $\zeta$ is the bulk viscosity of the electron liquid. In order to make this identification, we consider an arbitrary point $x_0$ along the wire and adjust the currents such that $j_E / T s_0(x_0) = I / e$. This relation ensures that $u(x_0) = v_d(x_0)$ and thus the electron liquid is fully equilibrated at this point. In order to maintain steady-state flow, a force must be applied to counteract the damping force~(\ref{eq:rhop2}). Denoting this force (per electron) by $f^{(\zeta)}$ and considering the small segment of the wire between the points $x_0 \pm \Delta x / 2$ containing $\Delta N$ electrons, we have $f^{(\zeta)} \Delta N = - \dot{\rho}_P^{(\zeta)} \Delta x$. The power dissipated is then given by $\Delta W = f^{(\zeta)} v_d \Delta N = - v_d \dot{\rho}_P^{(\zeta)} \Delta x$. Using Eq.~(\ref{eq:rhop2}) with $u = v_d$ and the continuity equation $\partial_x (n v_d) = 0$, we find that
\begin{equation}
\frac{dW}{dx} = \zeta (\partial_x v_d)^2.
\label{eq:diss}
\end{equation}
This equation represents the dissipation of a fluid due to its bulk viscosity~\cite{landau}, thus confirming the physical meaning of $\zeta$.

We now consider the effects of contacts on the transport properties of the wire. The leads control the distribution function of the excitations entering the wire. For example, the temperature of the left lead coincides with the temperature of the right-moving electrons at $x = 0$. The latter is not given by $T(0)$, but rather by $T(0)/(1-u/v)$. Therefore, there is a mismatch between the temperature of the leads and the temperature of the bosons at the end of the wire. This effect is fully accounted for by the contact thermal resistance~$R_T^{(0)}$. Adding the temperature drop $R_T^{(0)} j_E$ due to the contacts and $-\int \partial_x T dx$ [using Eq.~(\ref{eq:thermgrad})] gives
\begin{equation}
\Delta T = R_T^{(0)} j_E + \int_0^L \! dx \left[ \frac{j_E}{\kappa} - \frac{T s_0 I}{\kappa e} + \frac{\zeta}{T s_0^2} \left( \partial_x \frac{1}{n} \right)^2 j_E \right],
\label{eq:deltaT}
\end{equation}
where $R_T^{(0)} = 6 \hbar / \pi T$~\cite{thermal}.

The electric current through a single conductance channel (appropriate for spinless electrons) is given by $I = e^2 V / h$. Accounting for electron backscattering, we have $I = e^2 V / h + e \dot{N}^R$, which may be derived by considering electron current conservation~\cite{spinless}. Solving for $V$ and substituting Eqs.~(\ref{eq:rhop1}) and (\ref{eq:dotnr}) into $\dot{N}^R = \int \dot{n}^R dx$, we obtain
\begin{equation}
V = R^{(0)} I - \frac{1}{e} \int_0^L \frac{s_0}{\kappa} \left(j_E - \frac{T s_0 I}{e} \right) dx,
\label{eq:conductancecorrection}
\end{equation}
where $R^{(0)} = h / e^2$.

Equations~(\ref{eq:deltaT}) and (\ref{eq:conductancecorrection}) relate $\{\Delta T, V \}$ to the currents $\{j_E,I\}$. The currents are independent of position and can be factored out of the integrals. The various transport coefficients will now be expressed as integrals of the density $n(x)$, the entropy per particle $s_0(x)$, $\kappa(x)$, and $\zeta(x)$ over the length of the system.

The thermal resistance is defined at zero current and is given by $R_T = \Delta T/ j_E$. Equation~(\ref{eq:deltaT}) with $I = 0$ gives
\begin{equation}
R_T = R_T^{(0)} + \int \frac{dx}{\kappa} + \frac{1}{T} \int \frac{\zeta}{s_0^2} \left( \partial_x \frac{1}{n} \right)^2 dx.
\label{eq:thermres}
\end{equation}
The Peltier coefficient is the ratio of energy current to electric current, i.e. $\Pi = j_E / I$, when the left and right leads are at the same temperature. Setting $\Delta T = 0$ in Eq.~(\ref{eq:deltaT}) and solving for the ratio of currents yields
\begin{equation}
\Pi = \frac{T}{e R_T} \int \frac{s_0}{\kappa} \ dx.
\label{eq:pi}
\end{equation}
Since $\Pi \propto u / v_d$, this coefficient is directly informed by the competition between the processes described by Eqs.~(\ref{eq:rhop1}) and (\ref{eq:rhop2}). The thermopower is defined to be $S = - V/\Delta T$ with $I = 0$ and is straightforwardly obtained by dividing Eq.~(\ref{eq:conductancecorrection}) by Eq.~(\ref{eq:deltaT}). We find that $S = \Pi / T$, in accordance with the Onsager relations~\cite{onsager}.

Of central interest in experimental realizations of quantum wires is the resistance, $R = V/I$, defined for $\Delta T = 0$. By substituting $j_E = \Pi I$ into Eq.~(\ref{eq:conductancecorrection}), we obtain
\begin{eqnarray}
R = R^{(0)} + \frac{T}{e^2} \left[ \int \frac{s_0^2}{\kappa} \ dx - \frac{1}{R_T} \left( \int \frac{s_0}{\kappa} \ dx \right)^2 \right].
\label{eq:resistance}
\end{eqnarray}
Expressions (\ref{eq:thermres}), (\ref{eq:pi}), and (\ref{eq:resistance}) give a complete description of the transport properties of an inhomogeneous LL at temperatures below $D$ and represent our primary result.

We now establish the conditions for which Eq.~(\ref{eq:resistance}) agrees with the corresponding result of hydrodynamics. For long wires, Eq.~(\ref{eq:resistance}) reduces to Eq.~(4) of Ref.~\cite{hydro} provided
\begin{equation}
\frac{\zeta v \tau}{\hbar} \left( \frac{\hbar v}{T d} \right)^2 \ll 1,
\label{eq:tempregime}
\end{equation}
and the variations of the density are small along the wire, i.e. $\Delta n \ll n$. The condition~(\ref{eq:tempregime}) is controlled by the competition between the long characteristic length scale of the inhomogeneities $d$ and the exponentially long timescale $\tau \sim e^{D/T}$ associated with electron backscattering. The breakdown of the hydrodynamic theory occurs, to logarithmic accuracy, at a temperature
\begin{equation}
T^\ast \simeq \frac{D}{2 \ln \left( n d \right)}.
\end{equation}
Thus, in the temperature range $T^\ast \ll T \ll D$, the transport theory of Ref.~\cite{hydro} based on classical hydrodynamics applies to a quantum degenerate system.

The temperature dependence of the resistance (\ref{eq:resistance}) represents an important prediction of our theory. Since $R - R^{(0)} \propto \tau^{-1} \propto e^{-D/T}$, the resistance increases monotonically with temperature. The opposite behavior is exhibited by corrections to the resistance arising from short range disorder~\cite{schulz,kanefisher,furusaki}. So far, resistances increasing with temperature have been seen in conductance measurements of quantum point contacts~\cite{thomas,berggren}, systems in which the effects of impurities are easier to eliminate. Observation of such behavior in long wires would represent a confirmation of our theory.

\emph{Discussion.}--  In this work, we have calculated the transport properties of a spinless quantum wire subject to long range disorder with arbitrary electron-electron interactions. This work is readily generalized to spinful and other multi-channel electron liquids by taking appropriate values of $R^{(0)}$, $R_T^{(0)}$, $s_0$, $\kappa$, and $\zeta$. For sufficiently high temperatures, the degenerate electron liquid can be described within the hydrodynamic theory of Ref.~\cite{hydro}. From the vantage point of this more microscopic theory, we find that tacit to the hydrodynamic result is the requirement that the spatial variations in the density are small.

\emph{Acknowledgements.}--  We are grateful to A. V. Andreev for discussions. The work was supported by the U.S. Department of Energy, Office of Science, Materials Sciences and Engineering Division.

\end{document}